\documentclass[twocolumn,tighten,longauthor]{myaastex62}
\usepackage{amsmath,amsfonts,amssymb,amsthm}
\shorttitle{Short pulse duration limits of optical SETI}
\shortauthors{Michael Hippke}
\begin{document}
\title{INTERSTELLAR COMMUNICATION. XI. SHORT PULSE DURATION LIMITS OF OPTICAL SETI}
\author[0000-0002-0794-6339]{Michael Hippke}
\affiliation{Sonneberg Observatory, Sternwartestr. 32, 96515 Sonneberg, Germany}
\email{michael@hippke.org}

\begin{abstract}
Previous and ongoing searches for extraterrestrial optical and infrared nanosecond laser pulses and narrow line-width continuous emissions have so far returned null results. At the commonly used observation cadence of $\sim 10^{-9}\,$s, sky-integrated starlight is a relevant noise source for large field-of-view surveys. This can be reduced with narrow bandwidth filters, multipixel detectors, or a shorter observation cadence. We examine the limits of short pulses set by the uncertainty principle, interstellar scattering, atmospheric scintillation, refraction, dispersion and receiver technology. We find that optimal laser pulses are time-bandwidth limited Gaussians with a duration of $\Delta t \approx\,10^{-12}$\,s at a wavelength $\lambda_{0}\approx1\,\mu$m, and a spectral width of $\Delta \lambda \approx 1.5\,$nm. Shorter pulses are too strongly affected through Earth's atmosphere. Given certain technological advances, survey speed can be increased by three orders of magnitude when moving from ns to ps pulses. Faster (and/or parallel) signal processing would allow for an all-sky-at-once survey of lasers targeted at Earth.\\
\end{abstract}

\section{Introduction}
Optical SETI typically assumes laser signals to be either pulsed or narrowband. Short pulses are usually assumed to be $\Delta t_{\rm min} \leq 10^{-9}\,$s \citep{Howard2001} or $10^{-14} \dots 10^{-12}\,$s \citep{2016SPIE.9908E..10M}. Narrowband continuous emissions are generally taken as $\Delta \lambda < 10^{-2} \,$nm \citep[e.g.,][]{2017AJ....153..251T} or even $\Delta \lambda < 10^{-12} \,$nm \citep[$<\,$Hz,][]{1993SPIE.1867...75K}.

The justification for the feasibility of optical SETI is that short pulses or narrowband emissions outshine their blended host star as the main noise source \citep[e.g.,][]{2004ApJ...613.1270H}. This is plausible, because a laser focused and received through 10\,m telescopes would outshine a G2V host star for laser pulses of kJ energy at ns cadence (or in a narrow spectral channel), independently of distance, and there exist MJ (PW) lasers on Earth \citep{2017ApPhB.123...42H}.

Such examples offer a credible use-case for observations of individual stars one by one. However, a survey of the entire sky could be much faster by observing a large field of view. Unfortunately, noise levels increase with the number of stars observed and quickly require implausibly high signal power (or multipixel detectors). In order to reduce the noise levels, one could ``know'' (guess) the correct wavelength and use a narrow filter in addition to short time cadence. This scenario is explored in paper 10 of this series.

Another option, which is explored in this paper, is to increase the time resolution. For one hemisphere, the total night sky radiance over a bandwidth of $1{,}000\,$nm is $\approx 10^{14}$ photons per second (section~\ref{sub:atmo_noise}), or $\approx10^5$ photons per ns. This is the background flux which competes with a laser signal in an all-sky survey, and would overpower the typically assumed plausible OSETI signals. To make the background noise small per cadence, one could reduce the sky coverage, bandwidth, or a combination of both. At an observation cadence of order picosecond, the all-sky noise flux would reduce to $\approx 100$ photons per cadence. As typical wide-field telescopes have fields-of-view of $5^{\circ}\times5^{\circ}$ ($\sim 1/1{,}000$ of the sky, section~\ref{sub:telescopes}), noise levels reduce to 0.1 photons per ps cadence, which is negligible.

In this paper, we determine the physical short end time limit of laser pulses, set by the time-bandwidth limit, barycentric corrections, interstellar scattering, atmospheric refraction, dispersion, and broadening, and receiver technology.

\section{Physical limits}
\label{sec:physical}

\subsection{Time-bandwidth limit}
\label{sub:timebandwidth}
The Heisenberg uncertainty principle (time-bandwidth limit) prohibits an arbitrary combination of infinitely short and narrow pulses: ``the more precisely the position is determined, the less precisely the momentum is known in this instant, and vice versa'' \citep{Heisenberg1927}, so that
$\Delta E \, \Delta t \geq \hbar/2$ where $\Delta E$ is the standard deviation of the particle energy, $\Delta t$ is the time it takes the expectation value to change by one standard deviation, and $\hbar$ is the reduced Planck constant. A photon pulse with a temporal width $\Delta t$ can therefore not be monochromatic, but has a spectrum. Both are related through a Fourier transform, and it can be shown that \citep{Griffiths2004,Rulli2005}

\begin{equation}
\label{tmin}
\Delta t_{\rm min} \geq K \frac{\lambda_0^2}{\Delta \lambda\,c}
\end{equation}

where $\lambda_0$ is the central wavelength, $\Delta \lambda$ is the width of the spectrum (FWHM), $c$ is the speed of light and $K\sim0.441$ for a Gaussian pulse shape. For example, a near-infrared (NIR) laser pulse ($\lambda_0=1\,\mu$m) with a 5\,\% bandwidth ($\Delta \lambda=50\,$nm) has a minimum width of $\Delta t_{\rm min} \sim 29\times10^{-15}\,$s, or $\sim\,29\,$fs (Tables~\ref{tab:prefixes}, \ref{tab:time-bandwidth}).

\begin{figure}
\includegraphics[width=\linewidth]{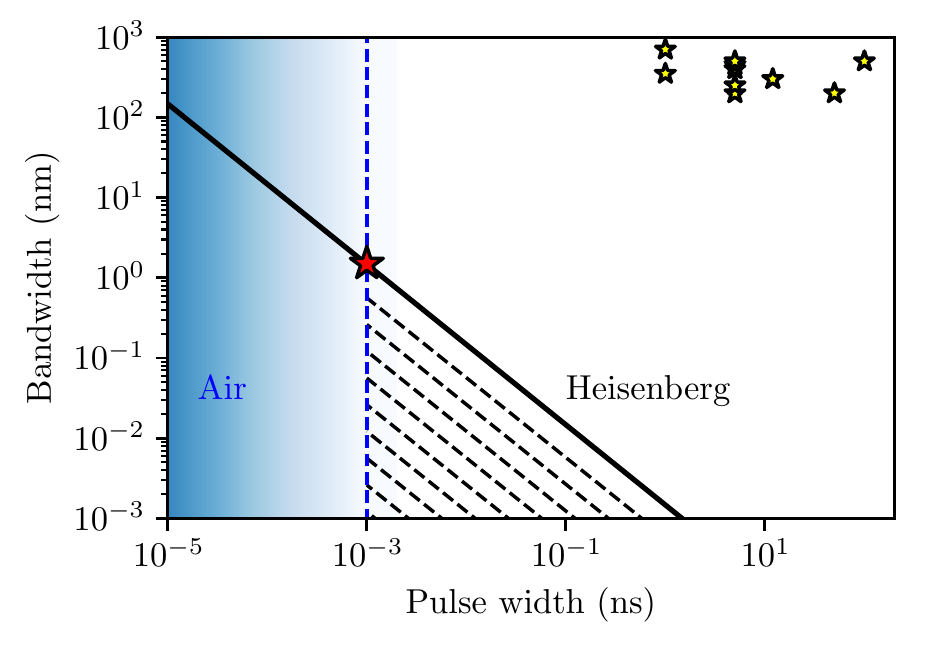}
\caption{\label{fig:oseti_heisenberg}Parameter space of a Gaussian pulse with a minimum time-bandwidth product for $\lambda=\,\mu$m (black line). Atmospheric effects (section~\ref{sec:atmo}) are shown in blue. Small star symbols represent past and current OSETI experiments (section~\ref{sub:previous_searches}), which are $\sim3$ orders of magnitude from the time-bandwidth limit. The red star symbol indicates the OSETI sweet spot identified in this paper.}
\end{figure}

As an example, \citet{2017AJ....153..251T} state that ``Laser lines as narrow as 1 Hz are already in use on Earth''. While this is a true statement, a Hz bandwidth laser pulse at $\lambda=1\,\mu$m has a pulse duration $\gtrapprox 0.4\,$s, so that a choice between ``broadband'' short pulses and ``continous'' wave narrowband emission is required. The allowed parameter space for pulses is shown in Figure~\ref{fig:oseti_heisenberg}. There is no obvious distinction where one regimes begins and the other ends. To comply with the literature, we will refer to ``pulses'' for signals with $\Delta t_{\rm min}<\,\mu$s.

\begin{table}
\center
\caption{Unit prefixes}
\label{tab:prefixes}
\begin{tabular}{ccc}
\hline
Text  & Symbol & Power     \\
\hline
milli & m     & $10^{-3}$  \\
micro & $\mu$ & $10^{-6}$  \\
nano  & n     & $10^{-9}$  \\
pico  & p     & $10^{-12}$ \\
femto & f     & $10^{-15}$ \\
\hline
\end{tabular}
\end{table}

\begin{table}
\center
\caption{Examples for time-bandwidth limited Gaussian pulses at $\lambda_0=1\,\mu$m}
\label{tab:time-bandwidth}
\begin{tabular}{cccr}
\hline
$\Delta \lambda$ (nm) & $\Delta \lambda / \lambda_0$ & $\Delta t_{\rm min}$ & Comment \\
\hline
50   & 0.05              & 29\,fs   & Shortest plausible pulse \\
1.5  & $2\times10^{-3}$ & 1\,ps   & Atmospheric limit \\
0.01 & $1\times10^{-5}$ & 0.15\,ns  & Normal spectroscopy \\
$10^{-6}$ & $1\times10^{-9}$ & $1.5\,\mu$s & Extreme spectroscopy\\
$3\times10^{-12}$ & $3\times10^{-15}$ & 0.4\,s & Hz bandwidth \\
\hline
\end{tabular}
\end{table}

\subsection{Minimum laser linewidth}
\label{sub:schawlow}
The minimum laser linewidth is also a function of power, which was known even before the first laser was experimentally demonstrated. The fundamental (quantum) limit for the linewidth of a laser is \citep{Schawlow1958}

\begin{equation}
\Delta f_{\rm laser} = \frac{2 \pi h f (\Delta f_\mathrm{c})^2}{P_\mathrm{out}},
\end{equation}

where $\Delta f_{\rm laser}$ is the half width at half-maximum linewidth of the laser, $\Delta f_\mathrm{c}$ is the half width of the resonances of the laser resonator and $P_\mathrm{out}$ is the laser output power. For OSETI, power levels will be large ($\gg\,$W) so that $\Delta f_{\rm laser} \ll \,$Hz (i.e., $\ll 10^{-12}\,$nm), and Schawlow-Townes produce no relevant limit.

\begin{figure*}
\includegraphics[width=.5\linewidth]{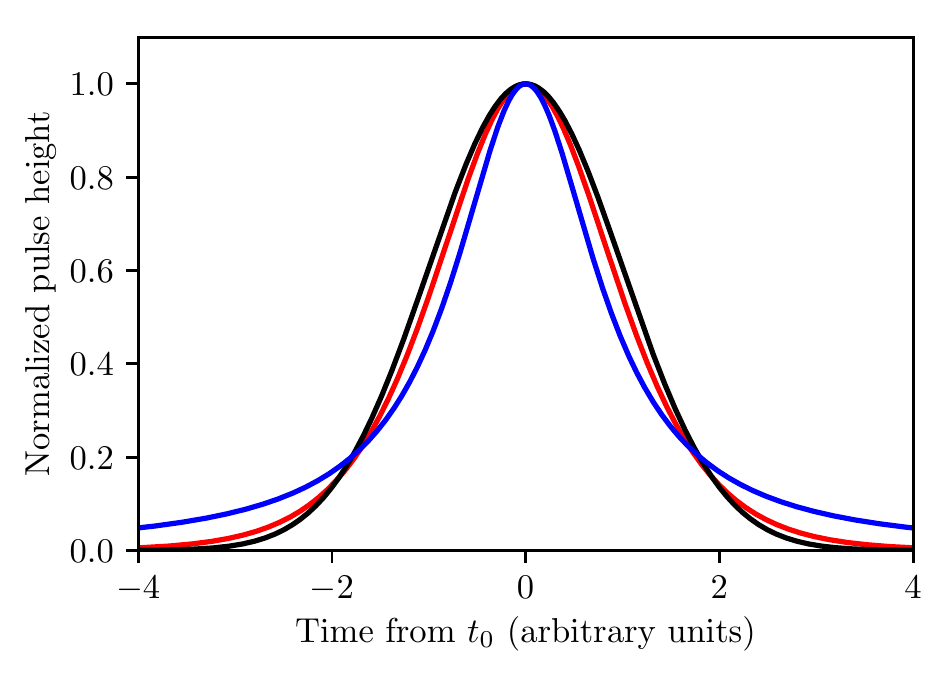}
\includegraphics[width=.5\linewidth]{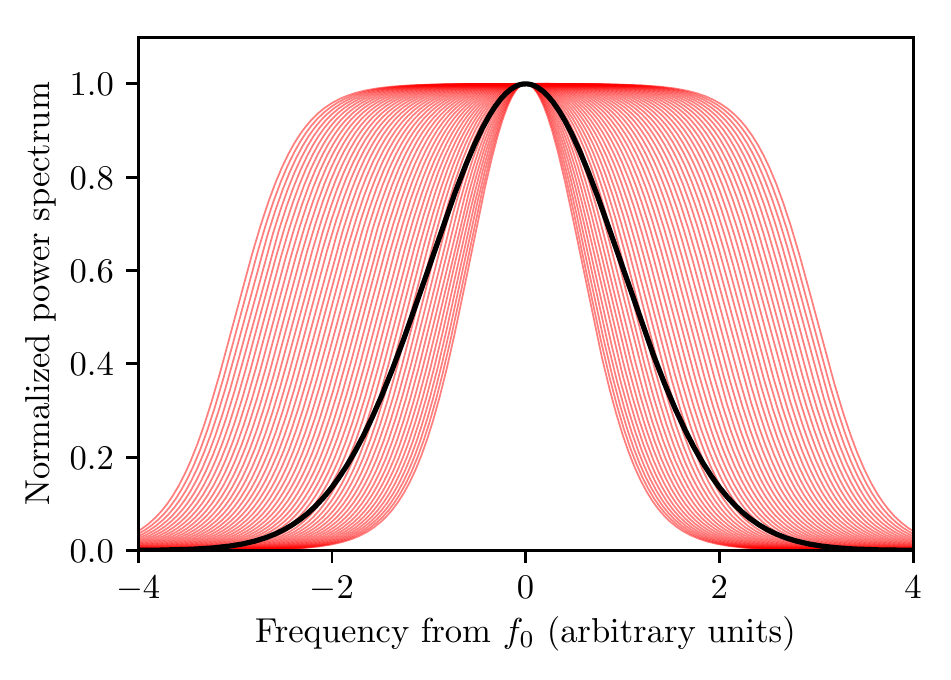}
\caption{\label{fig:sech2}Left: Comparison of pulse shapes with approximately the same energy integrals. Gaussian (black), sech$^2$ (red), Lorentzian (blue). Right: Spectrum of a Gaussian (black) and a sech$^2$ pulse for $0.5<\alpha<3$ (red).}
\end{figure*}

\subsection{Laser pulse shape}
\label{sub:pulse_shape}
Laser generation can produce pulses of various shapes in the time and frequency domains \citep{Smith1970,Hao2013}. Typically, the the temporal intensity of short laser pulses is approximated with a Gaussian

\begin{equation}
P(t)=P_{\rm P} \, {\rm exp} \left(-4\ln2 \left(\frac{t}{\tau}\right)^2\right)
\end{equation}

where $\tau$ is the pulse duration and $P_{\rm P}$ is the peak power. A minor discrepancy with reality is that the tails of a Gaussian function never actually reach zero (but the laser power does). Due to laser production technicalities, pulses from mode-locked lasers often have a temporal shape better described with a squared hyperbolic secant (sech$^2$) function \citep{Lazaridis1995}

\begin{equation}
P(t)= \frac{P_{\rm P}}{\mathrm{cosh}^2\left( \frac{t}{\tau} \right)}.
\end{equation}

For pulses of the same amplitude and energy, a Gaussian has slightly weaker wings than the sech$^2$-shaped pulse (Figure~\ref{fig:sech2}, left). For same energy same full-width at half-mean (FWHM) pulses, the Gaussian's peak power is higher, $\sim0.94$ the pulse energy divided by the FWHM pulse duration compared to $\approx0.88$ for the sech$^2$ pulse. The time-bandwidth product of a Gaussian is $K\sim0.441$ and $K\sim0.315$ for the sech$^2$ pulse, which allows for slightly shorter (FWHM) time-bandwidth limited sech$^2$ pulses.

The intensity spectrum of a Gaussian pulse is also a Gaussian. For sech$^2$ pulses, the spectrum depends on its phase-amplitude-coupling factor $\alpha$, and the normalized power spectrum is \citep{Lazaridis1995}

\begin{equation}
E(\omega) = \frac{{\rm sech} \left( \frac{\pi}{2} (\omega \tau + \alpha)\right) {\rm sech} \left( \frac{\pi}{2} (\omega \tau - \alpha) \right)  }{{\rm sech}^2(\pi \alpha / 2)}.
\end{equation}

A variety of shapes for $0.5<\alpha<3$ is shown in Figure~\ref{fig:sech2} (right), ranging from an approximately Gaussian to a top-hat function. For larger values of $\alpha$, the spectrum approximates a rectangle, which might be favorable given finite sized spectral windows of atmospheric transparency and low noise (see paper 10 of this series and Figure~\ref{fig:oseti_grid}).

A variety of other pulse shapes exists for other laser technologies, such as the Lorentzian shape (Figure~\ref{fig:sech2}, left). The observation of an alien laser pulse shape will tell us a lot about the technology used for its production. The following calculations in this paper will assume the most common Gaussian form.

\section{Interstellar effects}
\label{sec:interstellar}
Interstellar extinction is the absorption and scattering of photons by dust and gas. Absorption, mainly as a function of wavelength and distance, is discussed in paper 10 of this series. Relevant for the short pulse case is interstellar plasma scattering, which produces a temporal delay and pulse broadening in time and frequency.

\subsection{Dispersion}
\label{sub:interstellar_dispersion}
Dispersive time delays from interstellar plasma are well-known from pulsar studies and follow the relation \citep{1993ApJ...411..674T}

\begin{equation}
t_d \sim 4.15\,{\rm DM}\,f^{-2}  \,\,\,\, {\rm (ms)}
\end{equation}

valid for pulses $\lambda>\mu$m \citep{2011AcAau..68..366S}, where $f$ is the frequency in GHz and DM is the dispersion measure in units of pc\,cm$^{-3}$. The dispersion law corresponds to cold plasma and is valid where the photon frequency is much larger than the plasma frequency.

A typical dispersion measure is $\rm{DM}=1\,{\rm pc\,cm}^{-3}$ over 100\,pc in the solar neighborhood, ${\rm DM}=30\, {\rm pc\,cm}^{-3}$ over kpc distances and $\rm{DM}=100\,{\rm pc\,cm}^{-3}$ over kpc towards the galactic center \citep{2017ApJ...835...29Y}. For GHz frequencies and ${\rm DM}=30\,{\rm pc\,cm}^{-3}$, the delay is $t_d\approx0.1\,$s and decreases to $\approx1\,$ps at $\lambda=\,\mu$m.

Dispersive delays can approximately be removed by shifting the time-series of many narrow frequency channels with an estimated amount of DM for the source. This correction is never perfect because of the finite number of individual channels and time sampling, and the imperfect knowledge of the true (time variable) DM \citep{alder2012radio}. In a typical OSETI observation, where the receiver is monochromatic (e.g., a photomultiplier, PMT), de-dispersion is not possible. Even in case a fast spectrograph would be used, the correction can only be applied post-detection, as the DM is not known a priori. Overall, the time delay is of order ps or less, and changes little ($<1\,$\%, i.e. $\lesssim10\,$fs) over time.

\subsection{Scatter broadening}
\label{sub:interstellar_scatter_broadening}
Dust and protons in the interstellar medium scatter and absorb (then re-radiate) photons, reducing the pulse amplitude and leaving an exponential scattering tail \citep{2002astro.ph..7156C}. Such a tail could be modelled using an exponentially modified normal distribution \citep{Grushka1972}

\begin{equation}
{\displaystyle f(x)={\frac {\lambda }{2}}e^{{\frac {\lambda }{2}}(2\mu +\lambda \sigma ^{2}-2x)}\operatorname {erfc} \left({\frac {\mu +\lambda \sigma ^{2}-x}{{\sqrt {2}}\sigma }}\right)}
\end{equation}

where erfc is the complementary error function

\begin{equation}
{\begin{aligned}\operatorname {erfc}(x)&=1-\operatorname {erf}(x)\\&={\frac  {2}{{\sqrt  {\pi }}}}\int _{x}^{\infty }e^{{-t^{2}}}\,dt.\end{aligned}}
\end{equation}

\begin{figure}
\includegraphics[width=\linewidth]{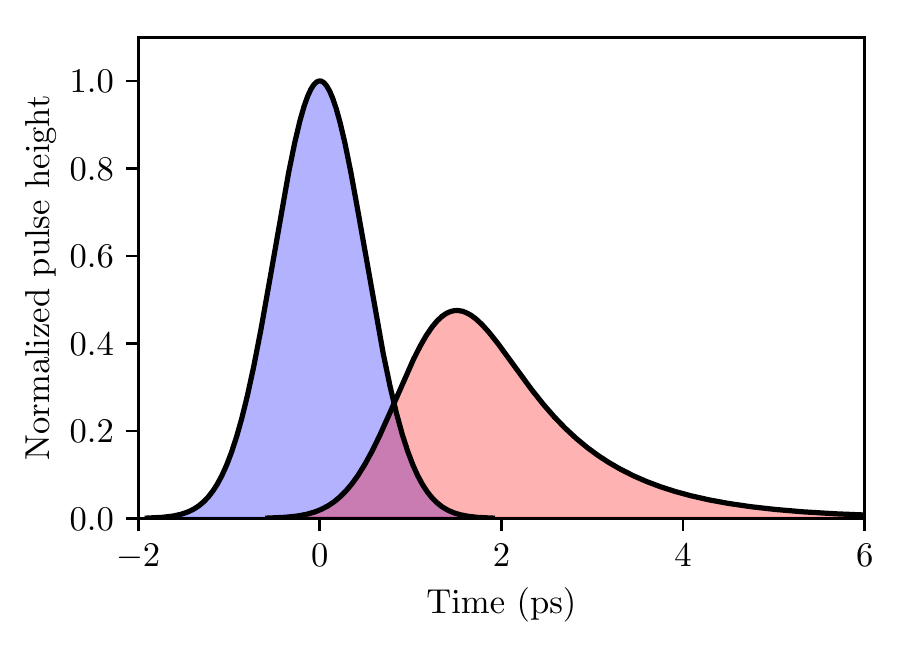}
\caption{\label{fig:tail}An original Gaussian pulse at FWHM=1\,ps (blue) is delayed and broadened (red). Pulse broadening is $<\,$fs in practice and exaggerated by $>100\times$ here for clarity.}
\end{figure}

This is illustrated in Figure~\ref{fig:tail}. On the one hand, it has been argued that the effect ``can be quite severe'' \citep{2000ASPC..213..545H,2001SPIE.4273..119H,2004ApJ...613.1270H}. On the other hand, conflicting statements exist, ``temporal dispersion (...) is entirely negligible at optical wavelengths.'' \citep{Howard2000}. Both arguments are based on calculations in \citet[][Appendix M.\,3]{Cordes2002} which state that a fraction of the pulse, $e^{-\tau}$ arrives unscattered, while $1-e^{-\tau}$ is broadened in time, with $\tau$ as the total optical depth. Therefore, scatter broadening is small for optical wavelengths out to 100\,pc, and in the IR out to kpc.

Frequency dependent scatter broadening is well known from pulsars, following a \citet{1941DoSSR..30..301K} turbulence model with a frequency scaling of $f^{-4}$. Pulsar scatter measurements closely follow this scaling over a large dispersion range, $2<{\rm DM}<1{,}000$ \citep{2015ApJ...804...23K}. However, the majority of sources show a flatter index over a frequency range between $0.1 \dots 1$\,GHz, with some down to $f^{-3}$ \citep{2017ApJ...846..104K,2017ApJ...835....2X}.

Using this scaling, the strongest scattering at GHz frequencies in our galaxy of $\approx1\,$s for DM=$1000\,$pc\,cm$^{-3}$ scales to $10^{-25}\,$s at $\lambda=\mu$m (or $10^{-17}\,$s at $f^{-3}$). This indicates that scattering tails at optical and IR wavelengths are irrelevant, because the time-bandwidth limit is stronger in all realistic cases.

This is in agreement with radio observations of the Crab pulsar which show pulses with high power (MJ) at $f=8.6\,$GHz which are unresolved at 2\,ns \citep{2003Natur.422..141H} time resolution and 0.4\,ns time resolution. Interstellar scattering models predict 0.05\,ns at $f=8.6\,$GHz for the Crab Nebula's DM of 56.7 \citep{2007ApJ...670..693H} at a distance of $1.9\pm0.1\,$kpc \citep{1973PASP...85..579T}.

At optical and IR wavelengths, practical measurements only exist into the ms regime. The scattering tail is unknown in practice. Limits from the Crab pulsar show no detectable scattering tail at UV and optical wavelengths for an optical millisecond pulse width and $E(B-V)=0.52$ \citep{2000ApJ...537..861S,Hinton2006,2007Ap&SS.308..595K,Lucarelli2008}.

These results indicate that the influence of the ISM is irrelevant for temporal pulse broadening at optical and IR wavelengths.

\subsection{Spectral broadening}
\label{sub:spectral_broadening}
Spectral broadening by the ISM and interplanetary medium is along most lines of sight well approximated as $\Delta f_{\rm broad}\sim0.1\,{\rm Hz}\,f^{-6/5}_{\rm GHz}$ \citep{1991ApJ...376..123C,2013ApJ...767...94S,2015aska.confE.116S}. For $\lambda=\mu$m ($f\sim10^{14}\,$Hz) we get $\Delta f_{\rm broad}\lesssim 10^{-8}\,$Hz, which is negligible.

\begin{figure*}
\includegraphics[width=\linewidth]{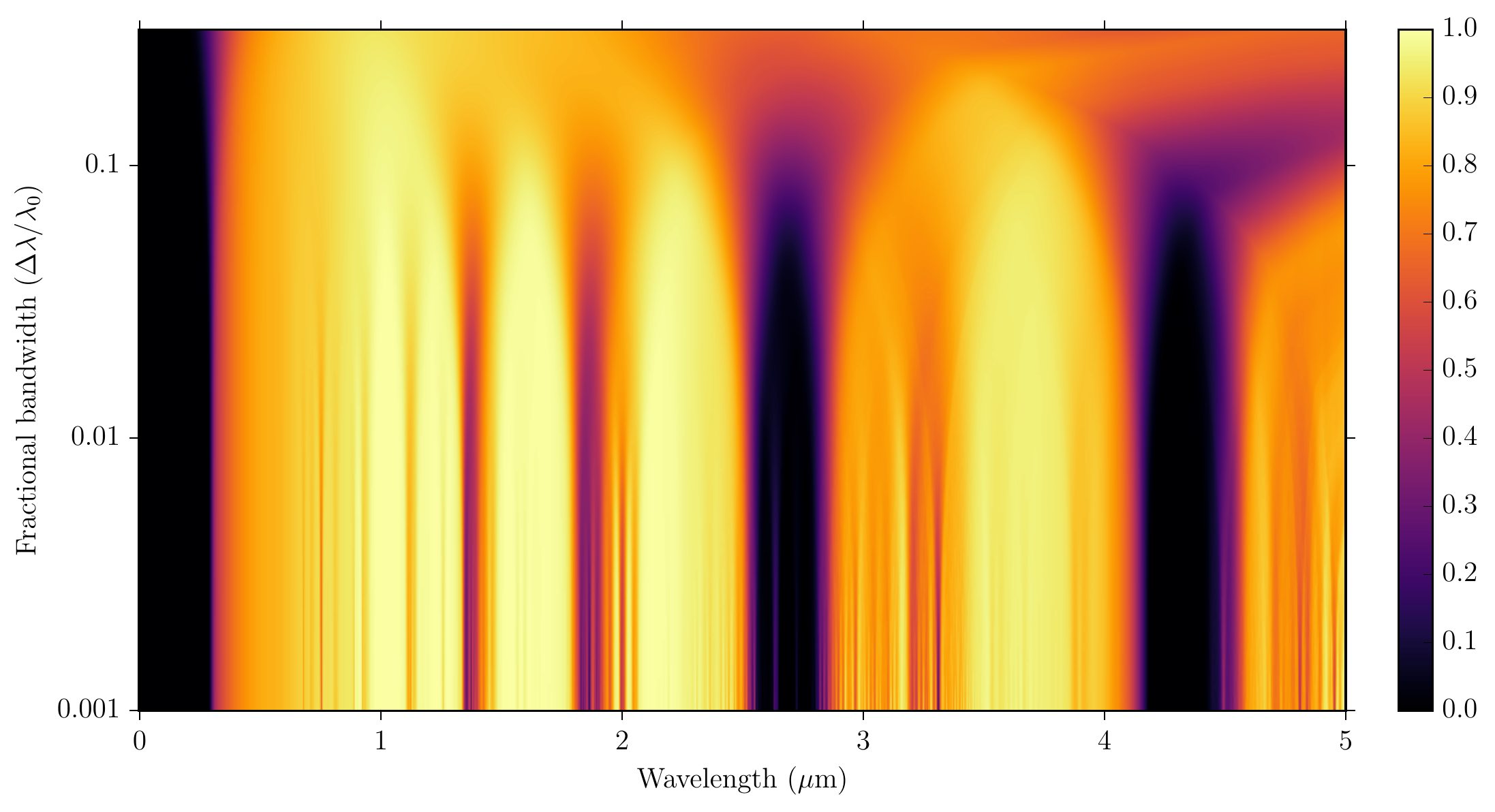}
\caption{\label{fig:oseti_grid}Atmospheric transparency as a function of wavelength and bandwidth, using a Gaussian convolution. Data from \citet{2012A&A...543A..92N,2013A&A...560A..91J} for VLT Cerro Paranal at an altitude of 2640\,m at median observing conditions with a precipitable water vapor of $2.5\,$mm at zenith angle.\\}
\end{figure*}

\section{Atmospheric effects}
\label{sec:atmo}
Dispersion and turbulence absorb and distort pulses travelling through Earth's atmosphere. Pulses are delayed compared to the speed of light in vacuum, broadened in length, and their temporal and spectral shapes are distorted. In addition, their amplitudes vary due to scintillation (``twinkling''), as discussed in paper 10 of this series and shown in Figure~\ref{fig:oseti_grid}.

The effects mainly depend on wavelength, initial pulse duration, and turbulence characteristics. The total delay and pulse broadening effects are of order ps under most conditions.

\begin{figure*}
\includegraphics[width=.5\linewidth]{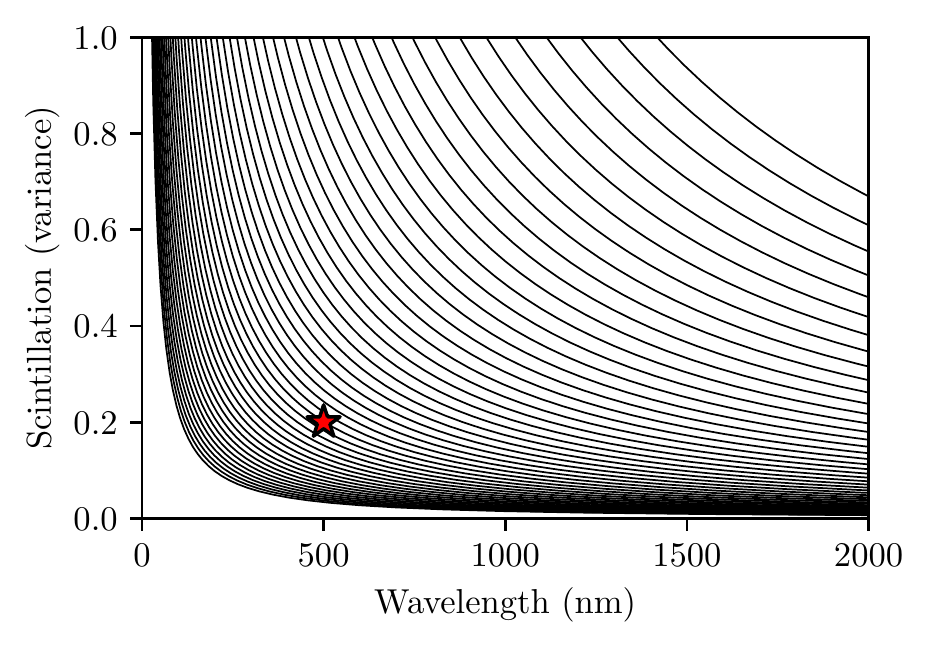}
\includegraphics[width=.5\linewidth]{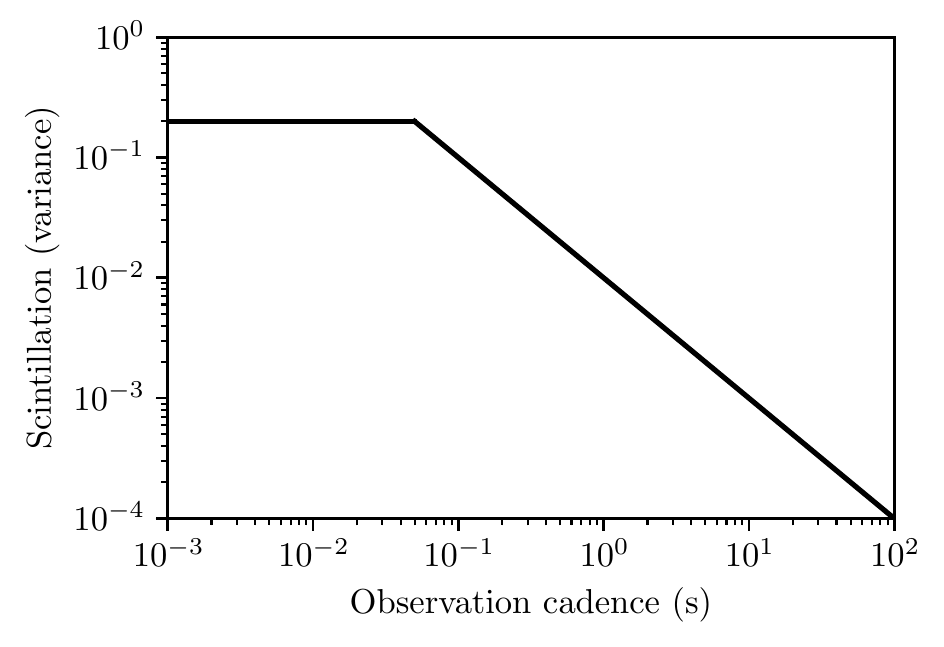}
\caption{\label{fig:scinti}Left: Estimated scintillation as a function of wavelength, with curves for medium to low turbulence ($10^{-15}<C^2<10^{-17}$) from space to ground. The red symbol shows the measurements taken with the 1\,m Jacobus Kapteyn Telescope on La Palma under typical conditions \citep{2015MNRAS.452.1707O}. Right: Scintillation as a function of observational cadence, scaled following Figure 10 in \citet{2015MNRAS.452.1707O}.}
\end{figure*}

\subsection{Atmospheric scintillation}
\label{sub:atmo_scinti}
Scintillation is a variation of the optical refractive index, caused by anomalous refraction through small-scale fluctuations in air density due to temperature gradients. It enlarges the point spread function of the telescope \citep{1995ApOpt..34.5461C} if not corrected for with adaptive optics \citep{1998aoat.book.....H}. The strength of scintillation is measured as the variance of the beam amplitude \citep[the Rytov variance,][]{1988ApOpt..27.2150A}

\begin{equation}
\sigma^2 = 1.23C^2 k^{7/6} H^{11/6}
\end{equation}

where $k = 2 \pi / \lambda$ is the wave number, H is the scale height of the atmospheric turbulence \citep[generally taken as $H\sim8{,}000\,$m,][]{2015MNRAS.452.1707O}, and $C^2$ is the structure constant for refractive-index fluctuations as a measure of the optical turbulence strength. Observed values are $C^2=1.7\times10^{-14}\,{\rm m}^{-2/3}$ at ground level and $C^2=2\times10^{-18}\,{\rm m}^{-2/3}$ at a height of 14\,km \citep{Coulman1988,Kopeika2001,Zilberman2001}. Common distinctions for turbulence levels are $10^{-13}$ (strong), $10^{-15}$ (average), and $10^{-17}$ (weak) \citep{goodman1985statistical,XiaomingZhu2002}. Turbulence is particularly low at Dome C in Antarctica, about $2 \dots 4 \times$ lower than at other observatories such as Cerro Tololo, Cerro Pach\'{o}n, La Palma, Mauna Kea, and Paranal \citep{2006PASP..118..924K,2015MNRAS.452.1707O}. Longer wavelengths experience a smaller variance with a factor of a few between optical and NIR wavelengths (Figure~\ref{fig:scinti}).

The median coherence time is typically $10 \dots 50\,$ms \citep{2006PASP..118..924K}, much longer than the relevant time scales of expected OSETI pulses (ns). In longer exposures, the scintillation noise will be reduced by temporal averaging \citep[for details, see][]{2015MNRAS.452.1707O}. A common approximation for the time averaging is given by \citet{1967AJ.....72..747Y}:

\begin{equation}
\sigma_{Y}^2 = 10^{-5}D^{-4/3}t^{-1}(\cos z)^{-3} \exp{\left(-2h/H\right)}
\end{equation}
where $D$ is the telescope aperture,
$z$ the zenith distance,
$t$ the observation cadence and
$h$ the altitude. For a 1\,s cadence of a meter-sized telescope under typical conditions, $\sigma_{Y}^2\sim10^{-5}$.

In practice, there is a cutoff point around the coherence time. The turning point is set by the amount of spatial averaging of the scintillation (the telescope aperture) and the wind velocity.

For the 1\,m Jacobus Kapteyn Telescope on La Palma under typical conditions, the knee is observed around $t\sim50\,$ms, below which scintillation changes to the short-exposure regime with amplitude variations of $\sim20\,$\% per 50\,ms cadence \citep{2015MNRAS.452.1707O}. Another station in Graz/Austria measured 200\,Hz (5\,ms) as the maximum frequency of the atmospheric fluctuations \citep{Prochazka2004,Kral2006}. Therefore, on ms and shorter timescales, scintillation will typically cause abrupt signal variations of $10\dots50\,$\% every $1\dots50$\,ms (Figure~\ref{fig:scinti}). Scintillation occurs as temporally autocorrelated (red) noise.

\subsection{Wavelength change from atmospheric refraction}
\label{sub:wave_ref}
While light travels at $c=299{,}792{,}458\,$m\,s$^{-1}$ in vacuum, its phase velocity $v$ is lower in a medium. In air, the refractive index is $n\approx1.00027717$, so that light travels at $v=c/n\approx299{,}709{,}388$\,m\,s$^{-1}$, causing a change in wavelength $\lambda$. The refractive index is mainly a function of pressure (altitude), water vapour, and carbon dioxide \citep{1966Metro...2...71E}. At $\lambda=1\,\mu$m, the wavelength change amounts to $\Delta \lambda \sim 0.27\,$nm, which is of the same order as the bandwidth of a time-bandwidth limited pulse. The transmitter or receiver may choose to adjust for this. Corrections are possible to better than $10^{-8}$, which makes the effect negligible \citep{1966Metro...2...71E}.

\subsection{Arrival time variations from refraction}
The delay in the arrival time of a pulse can be critical or irrelevant, depending on the magnitude of the effect and the search paradigm. For single pulses strong enough to ensure a significant detection, it is irrelevant whether these are recorded a little bit earlier or later - they are detected in any case. Single pulses, however, are of little value, because one can never be sure of their astrophysical origin. After such an initial detection, the search for a repeating pulse will be crucial. The most convenient repetition scheme would be a constant repetition, where the variations in arrival time are smaller than the instrumental measurement uncertainty. Therefore, it is interesting to determine the \textit{absolute} delay, as well as the \textit{change} in delay over time.

\subsubsection{Absolute refractive delay}
\label{sub:abs_ref_delay}
The approximate delay for the travel time from space to ground for a refractive index of $n\approx1.00027717$ and an air scale height of $H=8{,}000\,$m is $\approx7.5\,$ns at zenith ($\approx12\,$ns at $45^{\circ}$ elevation angle), with slight (0.5\,ns) variations from wavelength, temperature, pressure and humidity. During the course of an hour-long observing session, the elevation angle changes, causing arrival time variations of typically $1\dots10$\,ns. Searches for periodic signals require a correction for these effects. Detailed models for these parameters exist with experimental validation at the $<1\,$ps level \citep{Mendes2004}. Atmospheric correction models leave residuals of a few ps over an observing season \citep{Wijaya2011}.

\begin{figure*}
\includegraphics[width=.5\linewidth]{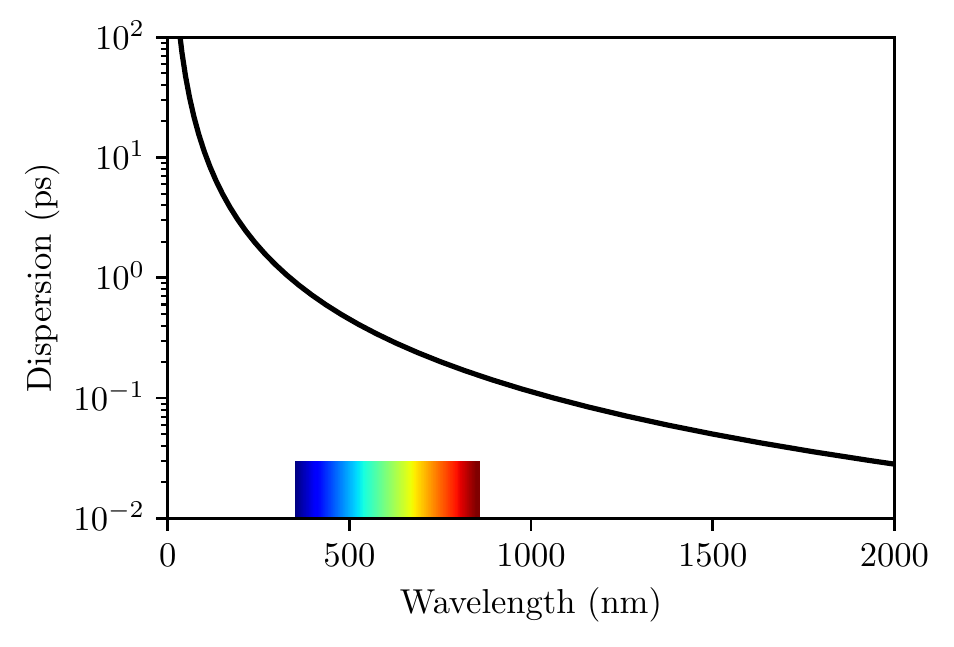}
\includegraphics[width=.5\linewidth]{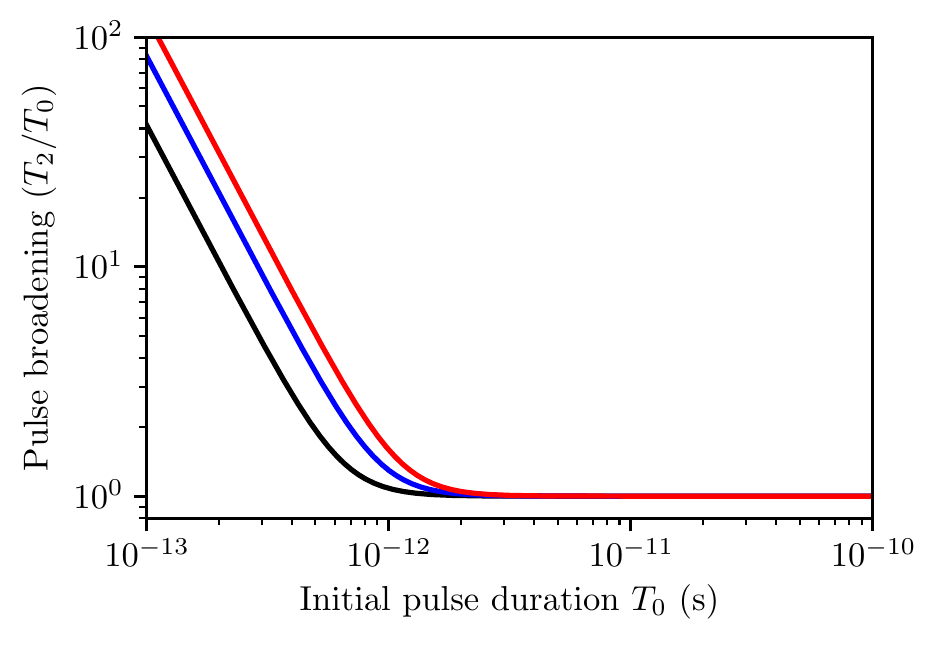}
\caption{\label{fig:chromatic_dispersive_pulse_broadening}Left: Dispersion for a pulse with $\Delta \lambda=100\,$nm and air distance of 20\,km as a function of wavelength. The dispersion is $<\,$ps for $\lambda > 300\,$nm. Right: Dispersive pulse broadening as a function of initial pulse duration. Colors show scenarios for travel distances from space to sea-level (red), mid ($2{,}500$\,m, blue) and high ($5{,}000$\,m, black) altitude. Initial pulses shorter than $10^{-12}\,$s are very sensitive to dispersion, because the square of the pulse duration is smaller than the group delay dispersion. The effect becomes negligible for pulses longer than $10^{-12}\,$s.}
\end{figure*}

\subsubsection{Variations of the refractive delay}
\label{sub:rel_ref_delay}
Fluctuations in the index of refraction of the air cause speed of light variations that cause a different time delay for a laser pulse \citep{Currie2014}. In practice, the effect has been measured with laser ranging observations to the Envisat satellite with the Graz Satellite Laser Ranging station \citep{Prochazka2004,Kral2006}. From an altitude of 500\,m to space and back, the atmospheric timing jitter was $0.6 \dots 1.0\,$ps, consistent with various models \citep{2007JGRB..112.6417H}. For a pulse traveling from space to Earth once (not twice), the effect is about half this value ($0.3 \dots 0.5\,$ps). At higher altitudes (e.g., $5{,}000\,$m), the jitter is likely smaller by a factor of a few, perhaps of order $0.1\,$ps.

\subsection{Pulse broadening through dispersion}
\label{sub:pulse_broad_disp}
Since there are no perfectly monochromatic pulses (section~\ref{sub:timebandwidth}), we have to consider the way in which a group of photons of different wavelengths travels through a medium such as Earth's atmosphere. A laser pulse can be treated as an envelope of wave amplitudes which travels through a medium with a group velocity

\begin{equation}
D_{\lambda} = {\rm GVD} \times \frac{-2 \pi c}{\lambda^2} \,\,\,\,(10^{-24}\,{\rm ps\,nm}^{-1}\,{\rm km}^{-1})
\end{equation}

where ${\rm GVD}\approx0.030036\,$fs$^2$\,mm$^{-1}$ is the estimated group velocity dispersion in air \citep{Ciddor1996}. For $\lambda=300\,$nm ($\lambda=\,\mu$m) and 8\,km of air, we get $D_{\lambda}\approx0.5\,$ps (0.05\,ps).

Dispersive pulse broadening also depends on the initial pulse duration $T_0$. Initial pulses $<\,$ps are very sensitive to dispersion, because the square of the pulse duration is smaller than the group delay dispersion. The effect becomes negligible for pulses $>$\,ps. For an originally unchirped Gaussian pulse with the duration $T_0$, the pulse duration is increased to \citep{boyd2013nonlinear}

\begin{equation}
T_2 = T_0 \sqrt{1+ \left( 4 \ln 2 \frac{D_2}{T_0^2} \right)^2 }
\end{equation}

where the group delay dispersion $D_2$ per unit length (in units of s$^2$\,m$^{-1}$) is the group velocity dispersion. For initial pulse durations $T_0=1$\,ps over 20\,km of air, $T_2\approx1.3\,$ps. The effect becomes negligible ($<1\,$\%) for pulses longer than a few ps (Figure~\ref{fig:chromatic_dispersive_pulse_broadening}).

\begin{figure}
\includegraphics[width=\linewidth]{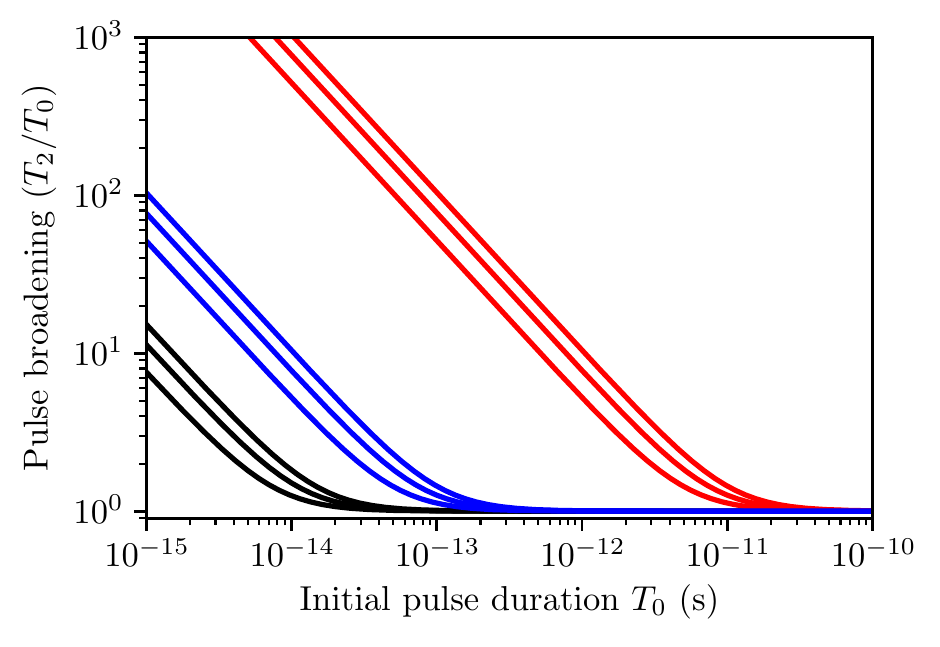}
\caption{\label{fig:turbi}Pulse broadening (as a factor of the initial pulse duration) due to turbulence, as a function of initial pulse length. Distance: Space to sea-level. Each series of curves shows wavelengths of 300\,nm (top), 500\,nm (middle) and $1{,}000\,$nm (bottom). Red: Strong turbulence ($C^2=10^{-14}\,$m, $L=10\,$m). Blue: medium turbulence ($C^2=10^{-16}\,$m, $L=10\,$m). Black: Quiet air ($C^2=10^{-17}\,$m, $L=1\,$m). Pulse broadening becomes negligible ($<1\,$\%) for $T_0>$\,ps in most cases and for $T_0>0.1$\,ns in all scenarios.}
\end{figure}

\subsection{Pulse broadening through turbulence}
\label{sub:pulse_broad_turbi}
Short optical signals passing through a turbulent medium such as Earth's atmosphere are temporally distorted, so that the received pulse has a longer duration. A semi-analytical model by \citet{1998ApOpt..37.7655Y} valid for the far-field \citep{TjinThamSjin1998} gives

\begin{equation}
T_2 = \sqrt{T^2_0 + 8 \alpha}
\end{equation}

where $z$ is the distance so that

\begin{equation}
\label{eqn:turbi}
\alpha = 0.322 \frac{\sigma^2 Q^{-5/6}}{(kc)^2},\,\,\, Q=\frac{z}{kL^2}.
\end{equation}

The model is valid for narrowband optical and IR pulses ($T_0>2 \times 10^{-14}\,$s). For the stratosphere and up, pulse broadening is very low ($<1\,$\%). Most relevant is the lowest layer, the troposphere, which contains $\approx90\,$\% of the air mass. Even in high turbulence, pulse broadening becomes negligible ($<1\,$\%) for $T_0>0.1$\,ns. In more quiet, photometric conditions, we expect pulse broadening to reduce to $<0.1\,$ps (Figure~\ref{fig:turbi}).

We note that \citet{2012OptCo.285.3169L} calculate pulse broadening to be more severe (about two orders of magnitude, at 10\,ps). The authors use a theoretical framework by \citet{Marcuse1981} for single-mode fibers which is likely not applicable to the atmosphere. Disagreements over which approximations describe reality best can be resolved by real-world measurements.

These theoretical estimates are in agreement with practical results. The atmospheric scattering in free-space laser communications ($\lambda=1.06\,\mu$m) is not relevant in practice for 10\,ps ($10^{-11}\,$s) pulses unless strong turbulence is present \citep{majumdar2010free}. The dependence of turbulent pulse broadening on wavelength is model-dependent and ranges between negligible \citet{1998ApOpt..37.7655Y} and a linear relation, with shorter wavelengths having stronger broadening \citep{Kelly1999}. In the latter model, temporal broadening is found to be very relevant for $10\dots30$\,fs pulses and becomes irrelevant for $>100\,$fs under all conditions (far field, near field, horizontal, vertical, strong turbulence) \citep{Kelly1999}.

\section{Barycentering}
\label{sec:bary}
Repeating signals may be periodic. The period can be constant or changing. A constant period can be constant in the ETI reference frame, or can be corrected to be constant (by ETI) for the Sun's or the Earth's reference frame. It is useful to understand the accuracy limits in order to perform the according periodicity tests for an assumed scenario.

It appears unlikely that a distant (e.g., kpc) periodic transmitter would be corrected for Earth's barycentric motion due to the dynamics of the solar system, such as nutation and precession of the Earth, over kyr of light time travel.

As we do not know the ETI's origin and thus reference frame (e.g., a transmitter in orbit around a star), we can only correct for the motion of known bodies. In our case, this is the Earth orbiting the sun.

The classical light travel time across the Earth's orbit (the R{\o}mer delay) has a magnitude of $\lesssim 498\,{\rm cos}\,\beta$ seconds with $\beta$ as the ecliptic latitude of the source, over the course of half a year; a rate of $<3.2\times10^{-5}$. Over the course of a one hour observing session, this amounts to $<0.12\,$s, and could smear any periodicity (of periodic signals) if uncorrected. Compared to this delay, other factors such as the Einstein and Shapiro delay are small ($<1.6\,$ms).

Simple barycentric correction codes offer corrections down to 0.24\,cm\,s$^{-1}$, or 29\,ns over an hour \citep{2014PASP..126..838W}. More complex implementations such as TEMPO2 \citep{2006MNRAS.369..655H} offer an accuracy of $\approx0.5\,$ns over 1\,hr and $\approx100\,$ns over years \citep{2006MNRAS.372.1549E}. These corrections have been validated against many pulsars with timing residuals of $\approx200$\,ns over 10\,yrs \citep{2010MNRAS.402.1027H}

\subsection{Spatial resolution}
\label{sub:periodicity}
In an all-sky at once observation (without high spatial resolution), barycentering corrections can not be made, because the ecliptic latitude of the source is unknown. With spatial resolution, the source location still has has uncertainties ($\delta \alpha$, $\delta \beta$) which give rise to periodic timing errors $\delta t_c$ resulting from incorrect barycentric correction \citep{lyne2006pulsar}

\begin{align}
    \delta t_c = &A\,\delta\alpha\,{\rm sin}(\omega t-\alpha)\, {\rm cos}\beta -\\
    &A\,\delta\beta\, {\rm cos}(\omega t-\alpha)\,{\rm sin} \beta \nonumber
\end{align}

which can amount to an absolute delay delta of a few seconds per degree on the sky. A source which is misidentified by one degree, and repeats periodically in the ETI reference frame, would change its period by $\lesssim 1\,$ms per hour in Earth's reference frame.

\subsection{Wavelength Doppler shift}
\label{sub:doppler}
Earth's orbit ($\lesssim30\,$km\,s$^{-1}$) and rotation ($\lesssim0.3\,$km\,s$^{-1}$) result in Doppler shifts in the received radiation spectrum. The corresponding shift over the course of half a year is $\Delta \lambda =\lambda_0 vc^{-1}\sim0.1\,$nm. This is smaller than the atmospheric limit for pulses ($\Delta \lambda \sim 1.5\,$nm for ps pulses, table~\ref{tab:time-bandwidth}). It is however larger by a factor of $\sim10$ than classical spectroscopy, and needs to be corrected in this case.

\section{Instrumental effects}
Before discussing telescope and detector choices, we review the technology used in previous OSETI experiments.

\begin{table*}
\center
\caption{Selection of previous pulsed OSETI detectors}
\label{tab:previous_obs}
\begin{tabular}{lcccr}
\hline
Obervatory & Cadence (ns) & $\lambda$ (nm) & Sensitivity ($\gamma$\,m$^{-2}\,$ns$^{-1}$) & Reference \\
\hline
MANIA 6\,m                     & 100 & $300\dots800$ & n/a & \citet{1993ASPC...47..381S} \\
Kingsley Columbus 0.25\,m      & 2  & $300\dots650$  & n/a & \citet{1995ASPC...74..387K} \\
OZ Australia 0.4\,m            & 1  & $300\dots650$  & n/a & \citet{2000ASPC..213..553B,2001SPIE.4273..144B} \\
Lick 1\,m Nickel               & 5  & $450\dots850$  & 51  & \citet{2001SPIE.4273..173W} \\
Harvard Oak Ridge 1.5\,m       & 5  & $450\dots650$  & 100 & \citet{2004ApJ...613.1270H} \\
Princeton Fitz Randolph 0.9\,m & 5  & $450\dots850$  & 80  & \citet{2004ApJ...613.1270H} \\
STACEE heliostats              & 12 & $300\dots600$  & 10  & \citet{2009AsBio...9..345H} \\
Leuschner 0.8\,m               & 5  & $300\dots700$  & 41  & \citet{2011SPIE.8152E..12K} \\
Harvard 1.8\,m                 & 5  & $300\dots800$  & 60  & \citet{2013PhDT.......161M} \\
Lick 1\,m NIROSETI             & 1  & $950\dots1650$ & 40  & \citet{2014SPIE.9147E..4KM} \\
Veritas 12\,m                  & 50 & $300\dots500$  & 1   & \citet{2016ApJ...818L..33A} \\
Boquete 0.5\,m                 & 5  & $350\dots600$  & 67  & \citet{2016ApJ...825L...5S} \\
\hline
\end{tabular}
\end{table*}

\begin{table*}
\center
\caption{Comparison of observation choices}
\label{tab:fluxes}
\begin{tabular}{ccccc}
\hline
Bandwidth (nm) & Cadence & G2V star at $d=100\,$pc & $5^{\circ}\times5^{\circ}$ & All-sky \\
$1{,}000$ & s  & $10^6$    & $10^{11}$ & $10^{14}$ \\
$1{,}000$ & ns & $10^{-3}$ & $100$     & $10^{5}$  \\
$1{,}000$ & ps & $10^{-6}$ & $0.1$     & $100$     \\
$1$       & ps & $10^{-9}$ & $10^{-4}$ & $0.1$     \\
\hline
\end{tabular}
\\Fluxes in photons per square meter.
\end{table*}

\subsection{Previous searches}
\label{sub:previous_searches}
Searches for continuous and pulsed laser signals began in the 1970s with efforts by \citet{1977SoSAO..19....5S,1997Ap&SS.252...51B}. Using small telescopes, fast PMTs entered the field in the 1990s \citep{1993SPIE.1867..178K,1995ASPC...74..387K} and were soon widely adopted \citep[e.g.,][]{2004IAUS..213..415W,2005AsBio...5..604S}. All modern searches for pulsed signals used PMTs with cadences of order ns (Table~\ref{tab:previous_obs}).

Searches with Cherenkov telescopes have been suggested by \citet{2001AsBio...1..489E,2005ICRC....5..387H}, and performed by e.g., \citet{2016ApJ...818L..33A}. Heliostats entered the field at about the same time \citep{1996APh.....5..353O,2009AsBio...9..345H}.

Only recently, quantum efficiency became sufficiently high to allow for useful nanosecond IR detectors, in discrete avalanche photodiodes (DAPDs) \citep{2014SPIE.9147E..0JW,2014SPIE.9147E..4KM,2016SPIE.9908E..10M}. While this project searches mainly for individual strong pulses, it uses a (weak) MHz pulsed lasers for tests.

Observations have mostly focused on FGK stars and individual objects, such as the anomalous star KIC~8462852 \citep{2016ApJ...825L...5S,2016ApJ...818L..33A}, and exoplanet host stars Trappist-1, GJ 422 and Wolf 1061 \citep{2018AAS...23110401W}.

The Harvard all-sky survey used a 1.8\,m telescopes with 16 PMTs, each with 64 pixels, scanning a field-of-view of $1.6^{\circ}\times0.2^{\circ}$ ($4\times10^{-6}$ of one sky hemisphere). Following Table~\ref{tab:fluxes}, the expected broadband flux per pixel is $10^{-3}$ per 5\,ns cadence \citet{2013PhDT.......161M}. This shows that broadband large field-of-view observations are possible even at ``slow'' ns cadence, using higher spatial resolution.

Spectroscopic (continuous wave) searches have been suggested \citep{1993ASPC...47..373B} and performed \citep{2002PASP..114..416R,2017AJ....153..251T} for $\sim5{,}600$ FGKM stars with power thresholds between 3\,kW and 13\,MW in a wavelength range $364 < \lambda < 789$\,nm.

\subsection{Telescope}
\label{sub:telescopes}
Light travels very slowly at 30\,cm\,ns$^{-1}$. This is relevant at short cadence, where all light from a collector must arrive at the detector within the same cadence, otherwise temporal smearing occurs, reducing the signal amplitude, and thus sensitivity. We now examine three exemplary telescope designs relevant for OSETI.

\subsubsection{Classical telescopes}
\label{sub:classical_telescope}
In a parabolic telescope mirror, parallel rays are perfectly focused to a point (the mirror is free of spherical aberration), no matter where they strike the mirror. However, this is only the case for rays that are parallel to the axis of the parabola, i.e. in the center of the field of view. Rays entering at an angle of a nonzero field of view will suffer from coma. This and higher order aberrations can be reduced with correctors, at the expense of additional throughput losses and costs. For typical parabolic reflectors, coma and astigmatism are $<10\,\mu$m for a field of view of a few degrees. The corresponding light time travel difference is $<10\,$fs, i.e. negligible. For meter-sized telescopes, practical limits to the field of view are of order $5^{\circ}$ \citep{2009JRASC.103...54R}. The LSST will have a FOV of $5^{\circ}$ \citep{2004SPIE.5489..705C,2016SPIE.9906E..0QN}. These limits keep light time travel differences $<\,50\,$fs, which means that atmospheric effects are larger by at least an order of magnitude. Similarly, mirror surface roughness of typically $<\lambda/4$ results in negligible ($<\,$fs) arrival time variations.

In case of multiple coincidence detectors, a precise alignment is required which takes into account the light travel time. For example, a ps cadence corresponds to a distance of 0.3\,mm.

\subsubsection{Fresnel lenses}
\label{sub:fresnel}
\begin{figure}
\includegraphics[width=\linewidth]{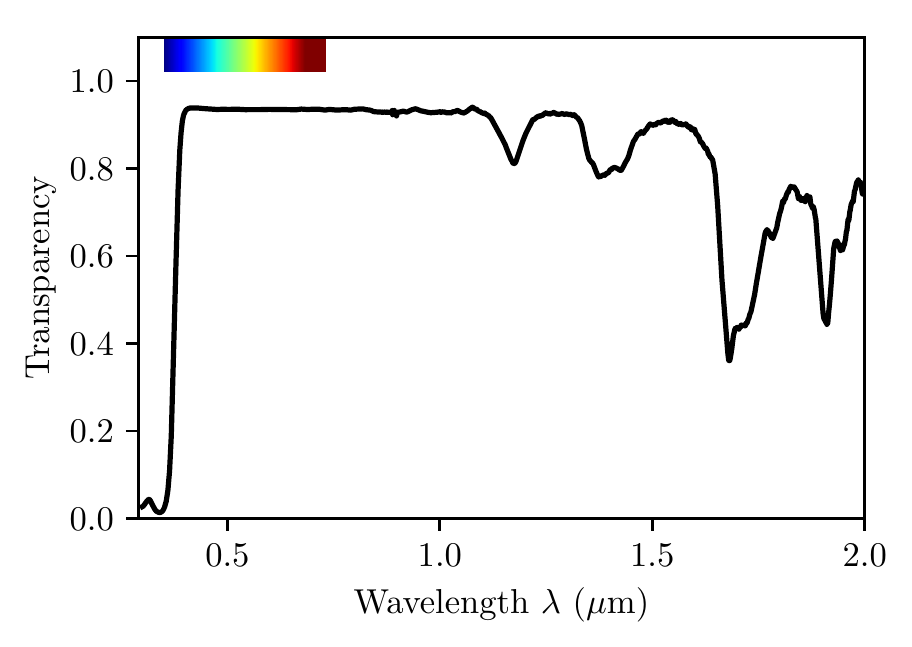}
\caption{\label{figure_oset_fresnel}Transparency as a function of wavelength for acrylic Fresnel lenses. Source: Thorlabs.}
\end{figure}

Fresnel lenses are thinner (typically mm), lighter and cheaper than classical lenses, as they divide the lens into a set of concentric annular sections. The finite size of these sections (typically mm) makes them much cheaper (hundreds of USD for a meter-sized aperture), but also of inferior optical quality, compared to classical lenses or parabolic mirrors. They can be made of plastics or acrylic, and their transparency can be of order unity for optical and IR wavelengths (Figure~\ref{figure_oset_fresnel}). Fresnel lenses are not diffraction limited, but can focus light to a mm spot size, so that their isocronicity error is a few ps \citep{2007ITNS...54..313C}. They have a large field of view, up to $30^{\circ}\times30^{\circ}$. It has been suggested to build a dedicated OSETI instrument in a dome, with an array of Fresnel lenses and detectors to cover the entire sky \citep{2013APS..APR.S2002C}. There are also plans to use Fresnel lenses for Cherenkov telescopes \citep{2002MmSAI..73.1211C,2008ICRC....3.1297M,2010arXiv1006.2266A}.

\subsubsection{Cherenkov Telescopes}
\label{sub:cherenkov}
Cherenkov Telescopes for the detection of very-high-energy gamma-ray photons offer large apertures for competitive prices. A Cherenkov telescope with a 12\,m aperture and $15^{\circ}$ field of view would exceed the \'{e}tendue of LSST by a factor of 10 \citep{2007APh....28...10V}. However, these telescopes are not isochronous. For example, the common Davies-Cotton telescope type with a total mirror area of order $10\,$m$^2$ located in 18 hexagonal facets of 0.78\,m has an optical time spread of $\Delta t\lesssim0.84\,$ns rms \citep{2013arXiv1307.3137M}. Its primary reflector forms a spherical structure towards the focal point, allowing for smaller aberrations off the optical axis compared to a parabolic design. Similarly, the H.E.S.S.-I telescope has a $\Delta t \sim 5\,$ns (rms $\sim1.4\,$ns) \citep{Akhperjanian2004,Schliesser2005}. Larger telescopes with the Davies-Cotton design would suffer from larger time spread \citep{Davies1957,2007APh....28...10V}. The effect could be reduced to $\lesssim0.3\,$ns by mounting the tessellated parabolic mirror facets staggered in depth \citep{2013APh....43..331D}.

Telescope designs such as the Schwarzschild-Couder \citep{1905MiGoe..10....1S} offer isochronous large unvigneted fields of view, up to $12^{\circ}$ \citep{Vassiliev2007,2008ICRC....3.1445V}. The isochronicity however is only valid on axis. Rays from the large field of view have a delay of $\sim0.3\,$ns per degree \citep[Figure 10 from][]{2007APh....28...10V}. In principle, this effect is correctable by delaying the pixels in the detector accordingly. If the detector has many ($10^4$) pixels, the field of view per pixel is small ($\sim$\,arcmin), so that the time smear effect per pixel is small ($\sim5\,$ps). There are plans to build detectors with many ($10^4$) pixels \citep{2011ExA....32..193A,2013APh....43....3A,2018arXiv180205715D}.

\subsubsection{Heliostats}
\label{sub:heliostats}
The problem of light time travel variations is particularly severe for extended light collectors such as heliostats, as used by STACEE, with 64 individual heliostats (each 37\,m$^2$) and a total collecting area of $2{,}300\,$m$^2$ \citep{2009AsBio...9..345H}. Synchronization was increased with separate detectors for groups of heliostats. Still, cadence was limited to 12\,ns.

\begin{table*}
\center
\caption{Detector technologies available for OSETI}
\label{tab:detector}
\begin{tabular}{ccccccc}
\hline
Detector &  Wavelength & Temp. (K) & QE (\%) & Time jitter $\Delta t$ (ps) & Dark count rate & Max count rate (MHz) \\
\hline
PMT                     & optical & 300 & 40 & 300 & 100 Hz  &    10 \\
PMT                     & IR      & 200 &  2 & 300 & 200 kHz &    10 \\
Si SPAD                 & optical & 250 & 65 & 400 &  25 Hz  &    10 \\
InGaAs SPAD             & IR      & 240 & 10 &  55 &  16 kHz &   100 \\
Frequency up-conversion & IR      & 300 &  2 &  40 &  20 kHz &    10 \\
SNSPD                   & IR      & 1.5 & 57 &  30 &   1 Hz  &  1000 \\
\hline
\end{tabular}
\\Data from \citet{Hadfield2009}
\end{table*}

\subsection{Detector}
\label{sub:detector}
An ideal detector would have a 100\,\% quantum efficiency, zero noise and dead time after a signal, and perfect time resolution. Commercial PMTs offer bandwidths of $\approx100\,$\%, quantum efficiencies of $\approx50\,$\%, dark rates of a few hundred Hz when cooled, reset times of $\approx3\,$ns and timing jitter of $\approx0.1\,$ns \citep{2010NIMPA.618..139A}. We show an overview of useful detector types for OSETI in Table~\ref{tab:detector}.

Superconducting nanowire single-photon detectors (SNSPDs) require cooling to a few K \citep{JinZhang2003} and provide ultrahigh counting rates exceeding 1\,GHz \citep{2008ApPhL..92x1112T}. Originally, their timing jitter was similar to PMTs, $\sim0.15$\,ns \citep{2001ApPhL..79..705G,2013NaPho...7..210M}. Latest improvements reduce timing jitter to very low values \citep[$<18$\,ps FWHM,][]{Shcheslavskiy2016}, at low intrinsic dark count rate \citep[$\ll$\,kHz,][]{2017SuScT..30kLT01G}, and short recovery times \citep[$<20\,$ns,][]{2017SuScT..30kLT01G}. Quantum efficiency is near unity in the optical and IR for single photons \citep[93\,\%,][]{2013NaPho...7..210M}.

Single Photon Avalanche Photodiodes (SPADs) offer a time resolution of $\sim50\,$ps, but a maximum count rate of 8\,MHz (125\,ns) \citep{2009JMOp...56..273B,2015SPIE.9504E..0CZ}.

Microwave Kinetic Inductance Detectors (MKIDs) provide large arrays at maximum count rates of $\sim10^3$ counts/pixel/s with $\mu$s timing \citep{2012RScI...83d4702M,2012OExpr..20.1503M}.

Timing resolutions of OSETI instruments have improved from 100\,ns \citep{1977SoSAO..19....5S} to 20\,ns \citep[][]{1997A&AT...13...13S} and finally 1\,ns \citep{2014SPIE.9147E..4KM}, a decrease by two orders of magnitude over 50 years.

\subsection{Electronics}
\label{sub:electronics}
Detectors can only be as fast as the readout electronics. Typical commercial sampling equipment works at GHz frequencies, and devices up to 10\,GHz (0.1\,ns) are common. Free-space optical communication is common at 0.56\,ns cadence (1.8\,GHz) \citep{Brandl2014,Ferraro2015}. The fastest commercially available signal processing oscilloscopes sample at 100\,GHz, or $10^{-11}\,$s \citep{2008Natur.456...81F,2012MeScT..23b5201F}. In the laboratory, petahertz optical oscilloscopes ($10^{-15}\,$s) have been demonstrated \citep{2013NaPho...7..958K}.

In practice, fast detectors such as SNSPDs can be sampled at$\sim10\,$ps time resolution as long as the count rate is $\lesssim 10\,$MHz (dead time 100\,ns), as demonstrated by \citet{Shcheslavskiy2016} using commercial equipment.

\section{Noise}

\subsection{Stochastic background}
\label{sub:atmo_noise}
Measurements of the night sky brightness are available for many observatory sites, such as
La Palma \citep[21.9\,mag\,arcsec$^{-2}$ in V-band,][]{1998NewAR..42..503B},
San Pedro Martir (Mexico) \citep[21.84,][]{2017PASP..129c5003P},
Calar Alto \citep[22.01,][]{2007PASP..119.1186S} and
Dome A in Antarctica \citep[23.4,][]{2012PASP..124..637S,2017AJ....154....6Y}.

A detailed model of the night sky spectrum (Cerro Paranal Advanced Sky Model) is consistent with these values\footnote{For the conversion of light-intensity units, see \citet[][Appendix]{1998NewAR..42..503B}} \citep{2012A&A...543A..92N,2013A&A...560A..91J}. This is also consistent with measurements using a large field-of-view telescope with a PMT, resulting in $2\times10^{12}$ $\gamma\,$s$^{-1}$\,sr$^{-1}$\,m$^{-2}$ for $300<\lambda<650\,$nm for dark sky regions such as Sculptor or Virgo, and about twice the value for bright regions such as Carina towards the galactic plane \citep{2011AdSpR..48.1017H}, in agreement with measurements from Namibia \citep{Preu2002} and La Palma \citep{mirzoyan1994measurement}.

Based on these counts, the total radiance for one sky hemisphere ($2.7\times10^{11}\,$arcsec$^2$) is $\approx 10^{14}\,\gamma\,$s$^{-1}\,$m$^{-2}$. With a 1\,nm filter centered at $\lambda=1.064\,\mu$m, the radiance is $<10^{11}\,\gamma\,$s$^{-1}\,$m$^{-2}$, or about half that value for telescope and receiver efficiencies of 50\,\%. This is still $\lesssim100\,\gamma\,$ns$^{-1}\,$m$^{-2}$. The all-sky background would be very small ($\lesssim 0.1$) at a picosecond observation cadence, or when observing a fraction of the sky.

In a typical wide-field telescopes with a fields-of-view of $5^{\circ}\times5^{\circ}$ ($\sim 1/1{,}000$ of the sky, section~\ref{sub:telescopes}), noise levels reduce to 0.1 photons per ps cadence for a bandwidth of $1{,}000\,$nm (Table~\ref{tab:fluxes}).

\subsection{Short natural astrophysical pulses}
A detailed study of astrophysical phenomena found that the shortest timescale of (known) natural signals in the optical appears to be of microsecond duration and longer \citep{2001SPIE.4273..153H}. If that is true, no optical ``RFI'' exists at ns or ps cadence. The discovery of such a source would be of great interest in any case.

Terrestrial interference comes from Air Cerenkov flashes produced by cosmic rays and $\gamma$-rays. These have typical durations of $\sim5\,$ns and can be distinguished by imageable tracks in multi-pixel detectors \citep{2001AsBio...1..489E}. Similarly, OSETI observations have been triggered by airplane positioning lights with $\mu$s durations. These signals can be flagged with a dome camera which detects bright moving objects. Such events are not periodic on the relevant time scales \citep{2013PhDT.......161M}.

\begin{figure}
\includegraphics[width=\linewidth]{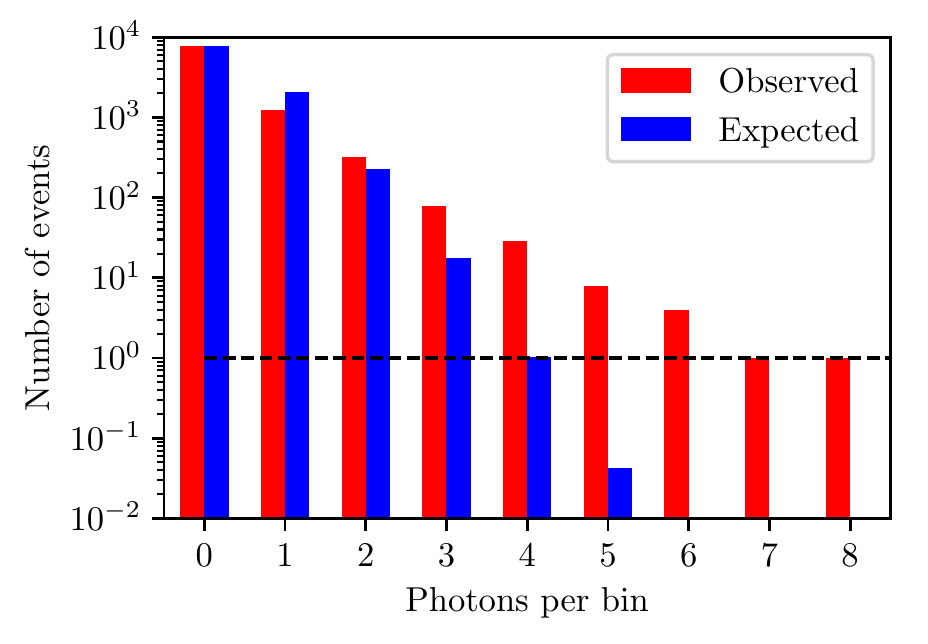}
\caption{\label{fig:bino}Received photons per timeslot in the lunar laser ranging experiment \citep[red,][]{2008PASP..120...20M} versus expected photons following a Poisson distribution (blue). To explain the 6, 7 and 8 photon events, the flux level has to be $6\times$ higher, which can not be explained with better atmospheric transparency which is of order 50\,\%. The difference can only be due to laser power variations and pointing quality fluctuations .}
\end{figure}

\subsection{Poissonian photon pileup}
Using beamsplitters and multiple coincidence detectors, the false positive rate can be reduced. In the absence of any true pulsed signal, the expected number of coincidences $R$ is \citep{2001SPIE.4273..173W,2002PhDT........13C}

\begin{equation}
R = \left( \frac{r}{n} \right)^n \, \eta_{\rm PMT}\, T^{n-1}\,
\end{equation}

where $n \geq 1$ is the number of detectors, $r$ is the number of photons per second, $T$ is the pulse width, and $\eta$ is the detector efficiency. For example, a rate of $10^6$ detected photons per second leads to $9\times10^5$ coincidences per hour with two detectors, and 133 per hour with 3 detectors for $\eta=1$. Using a more realistic $\eta=0.2$, it reduces to 27\,hr$^{-1}$. The downside of this false positive signal reduction by multiple coincidence detection is a loss in sensitivity to $<1/n$.

For the search paradigm of repeating signals, false positives (or a noise floor) are not critical, as these effects vanish with well-known search methods for periodicity, such as FFTs, periodograms and autocorrelation.

In practice, data might occur not to exhibit strict Poisson noise, as is observed in lunar laser ranging, which measures the distance between the Earth and the moon with the light travel time from Earth-based lasers bounced back by retro-reflectors placed on the lunar surface \citep{2008PASP..120...20M,2012CQGra..29r4005M}. These observations originally used few high power pulses \citep{1985ITGRS..23..385S,1998A&AS..130..235S,2013RPPh...76g6901M}, but recently moved towards lower pulse energies at much higher repetition frequency (80\,MHz) Nd:YAGs at $\lambda=1064\,$nm and $\lambda=532\,$nm with typical widths $\Delta \lambda\sim0.19\,$nm in combination with short (10\,ps) pulse durations \citep{2017CQGra..34x5008A}. Many weaker pulses have the advantage that their arrival time errors can be averaged, and a finite amount of monetary investment buys more \textit{average} laser power (at lower peak power). This concept has been proposed for OSETI by \citet{2013AsBio..13..521L}.

Lunar laser ranging observations have been found to exhibit a strong shot-to-shot variability ascribed to speckle structure and other scintillation (``seeing'') effects. At an average return rate of 0.23 photons per pulse, statistics would expect $<20$\,\% of the returns to be in multiple-photon bundles, and no events with more than four photons in a $10{,}000$-shot run. In practice \citep{2008PASP..120...20M}, $46\,$\% of the returning photons were in multiple-photon bundles, with up to 8 photons detected in one cadence (Figure~\ref{fig:bino}). For OSETI, the pointing quality is irrelevant, as the ETI laser would illuminate all of Earth equally. Atmospheric scintillation, however, may contribute $10\dots50\,$\% of flux variation, and may lead to higher than expected photon pile-up.

\begin{table*}
\center
\caption{Summary of effects on short pulses ($\lambda_0=1\,\mu$m, $\Delta \lambda = 1\,$nm)}
\label{tab:summary}
\begin{tabular}{lccc}
\hline
Component & Effect & Comment & Section \\
\toprule
Physical limits: & & & \ref{sec:physical} \\
-- Time-bandwidth limit & $\approx\,$ps & Choice based on $\lambda_0$ and $\Delta \lambda$ & \ref{sub:timebandwidth} \\
-- Schawlow-Townes & $\Delta f_{\rm laser} \ll \,$Hz & Irrelevant & \ref{sub:schawlow} \\
-- Pulse shape & $\sim10\,$\% of pulse duration & Irrelevant & \ref{sub:pulse_shape} \\

\hline
Interstellar: & & & \ref{sec:interstellar}\\
-- Dispersion & $\approx\,$ps & Constant delay; variation $<1\,$\% & \ref{sub:interstellar_dispersion} \\
-- Scatter broadening & $<\,$fs & Irrelevant & \ref{sub:interstellar_scatter_broadening} \\
-- Spectral broadening & $<10^{-8}$\,Hz & Irrelevant & \ref{sub:spectral_broadening} \\
\hline
Atmospheric: & & & \ref{sec:atmo} \\
-- Scintillation & $\lesssim20\,$\% variance & Typical frequency $\approx200\,$Hz & \ref{sub:atmo_scinti} \\
-- Wavelength change (refraction) & $\lesssim 0.3\,$nm & Correctable to $<10^{-4}\,$nm & \ref{sub:wave_ref} \\
-- Absolute refractive delay & $\lesssim 10\,$ns & Correctable to $\sim\,$ps & \ref{sub:abs_ref_delay} \\
-- Refractive delay variations & $\lesssim0.5\,$ps & Likely not correctable & \ref{sub:rel_ref_delay} \\
-- Pulse broadening (dispersion) & $T_2\lesssim 0.3\,$ps for $T_0=1\,$ps & For medium turbulence & \ref{sub:pulse_broad_disp} \\
-- Pulse broadening (turbulence) & $T_2\lesssim 0.1\,$ps for $T_0=1\,$ps & For medium turbulence & \ref{sub:pulse_broad_turbi} \\
\hline
Barycentering & $<0.12\,$s per hour & Correctable to $\lesssim 0.5\,$ns per hour & \ref{sec:bary} \\
-- Finite spatial resolution & $<\,$ms per degree per hour & Relevant for periodic signals & \ref{sub:periodicity} \\
-- Spectral Doppler shift & $<0.1\,$nm per 6 months & Correctable to $<10^{-4}\,$nm & \ref{sub:doppler} \\
\hline
Telescope & & & \ref{sub:telescopes} \\
-- Parabolic reflector & $<10\,$fs & Instrument alignment relevant & \ref{sub:classical_telescope} \\
-- Fresnel lenses & $\sim3\,$ps & & \ref{sub:fresnel} \\
-- Cherenkov telescopes & $\sim0.3$\,ns per degree & Correctable with multipixel detectors to $\sim5\,$ps & \ref{sub:cherenkov} \\
-- Heliostats & $\sim10\,$ns & & \ref{sub:heliostats} \\
\hline
Detector & $>30\,$ps & Depending on type, see Table~\ref{tab:detector} & \ref{sub:detector} \\
\hline
Electronics & $>10\,$ps & See text & \ref{sub:electronics} \\
\hline
\end{tabular}
\end{table*}

\section{Summary of results}
All relevant effects are summarized in Table~\ref{tab:summary}. Interstellar effects are irrelevant in all cases. Pulses are broadened by $\lesssim0.3\,$ps by dispersion and $\lesssim0.1\,$ps by turbulence in the atmosphere. Time of arrival variations are dominated by changes in refraction to $\lesssim0.5\,$ps. Most atmospheric effects are also a function of wavelength, and are factor of a few less severe for NIR compared to optical.

Given these atmospheric limits, pulses shorter than ps are not realistic for ground-based detectors. Such ps pulses have time-bandwidth limits of $\Delta \lambda \sim\,$nm at $\lambda_0=1\,\mu$m. This corresponds to a fractional bandwidth $\Delta \lambda / \lambda_0 \sim 0.001$. As can be seen in Figure~\ref{fig:oseti_grid}, there are windows of atmospheric transparency (discussed in detail in paper 10 of this series) of order unity even for large bandwidths, $\Delta \lambda / \lambda_0 \lesssim 0.1$.

Periodic signals are additionally affected by barycentering issues. While spectral Doppler shifts are correctable, current timing codes are limited to an accuracy of order ns per hour. This accuracy may be improved in the future with more detailed models and calibrations by many pulsars. For repeating ps pulses which occur e.g. at kHz repetition over a few seconds, current corrections are sufficient. Over longer times (years), many subtle adjustments would need to be made. For example, continental drifts are a few cm per year, which translates to $\gtrsim 100\,\,$ps\,yr$^{-1}$ (or $\gtrsim 0.3\,\,$ps\,day$^{-1}$). It appears sensible to restrict searches of periodic ps pulses to short time durations of order minutes.

\section{Discussion}

\subsection{Short signals through spectroscopy}
There is one additional method for OSETI, which is widely ignored in the literature. It is based on spectra which are Fourier transformed to search for periodic modulations. In this section, we explain the method and discuss its advantages and issues.

The spectral modulation of coherently separated laser pulses was first noted by \citet{1992ApOpt..31.3383C} and subsequently studied in \citet{2010ApJ...715..589B,2010A&A...511L...6B,2012AJ....144..181B} with the motivation to apply the method to astronomical data. Astronomical spectra are sampled at equal wavelength intervals, $\lambda = c/f$ where $\Delta \lambda={\rm const}$. To search for signals with a constant temporal period, these spectra must be converted to equal frequency intervals $f=c/\lambda$ with $\Delta f ={\rm const}$. With constant frequency intervals, a Spectral Fourier transform (SFFT) can me made which produces power for signals with a constant temporal period, such as repeating laser pulses.

Searches for such periodic temporal modulations were reported for SDSS spectra of individual stars \citep{2016PASP..128k4201B,2017JApA...38...23B} as well as galaxies \citep{2013ApJ...774..142B}. Detections were claimed with repetition frequencies (not pulse durations) of $\approx 10^{-13}\,$s.

For repetition frequencies of $\approx 10^{-13}\,$s, the pulse duration must be shorter than this value. Due to atmospheric pulse broadening, this appears impossible, and the results are likely instrumental artifacts. In general, however, the method appears to be useful for signals with longer repetition frequencies, but would require independent confirmation with another method in case a signal would be detected.

SFFTs are not sensitive to the pulse length, but instead to the duration between pulses ($\rho)$, where $1/\rho$ is the pulse repetition rate. SFFTs' sensitivity region is constrained by the spectral resolution, typically $2.5\times10^{-15} \dots 4\times10^{-12}\,$s for optical spectra with $R\approx10{,}000$ and can be increased to longer spacings ($10^{-11}\,$s) with higher resolution ($R=100{,}000$) spectrographs. The spectrograph ``ESPRESSO'' is expected to deliver a resolution of $R=200{,}000$ between $380<\lambda<780$\,nm at 5\,\% efficiency, a linewidth of 0.001\,nm \citep{2017arXiv171105250G}.

As the method of spectral Fourier transforms (SFFT) is not intuitive and requires careful implementation, we provide an open-source \texttt{Python} solution for future tests\footnote{\url{http://github.com/hippke/laserpulses}}.
The sensitivity of SFFTs is limited to about $10^{-5}$ of the stellar flux in SDSS spectra, if the star is blended with the hypothetical laser source. For a $L=L_{\odot}$ star and a competing laser with $D_{\rm t}=D_{\rm r}=10\,$m at $\lambda=1\,\mu$m, this SNR can be achieved with an average laser power of 3.5\,MW independently of distance. For comparison, military laser weapons are being developed with $\approx0.2\,$MW power. For faster survey speeds, multiple spectra can be obtained in parallel, as was done by SDSS.

Overall, SFFTs are appealing for signals with very high repetition rates. They are limited to $10^{-10}\,$\,s$\,<\rho<10^{-12}\,$s, with limits set by the resolution of spectroscopy and atmospheric pulse broadening.

\subsection{Data volume of a ps cadence broadband all-sky survey}
We now estimate the data volume and computing power requirements for observations at high (ps) cadence. For an all-sky survey, the stochastic sky-integrated background of starlight and atmospheric noise is $\approx10^{14}\,\gamma\,$s$^{-1}$\,sr$^{-1}$\,m$^{-2}$ for optical and NIR wavelengths. Sampled at ps cadence, at least 100 channels (spatial or color) would be required for less than unity flux per cadence, in order to push noise to less than one photon per cadence.

Every cadence and channel could be sampled with e.g., one byte of data to allow for 256 distinct values. The data stream is then 100\,GB\,s$^{-1}$. For comparison, the Breakthrough Listen Radio Data Recorder \citep{2018PASP..130d4502M} saves 24 GB\,s$^{-1}$ of data to disk. Their computing facilities also allow for real-time de-dispersion and pulse search. It appears that an all-sky optical and NIR survey at ps cadence would not require implausible computing requirements.

\section{Conclusion}
We have examined the influence of interstellar and atmospheric effects on short pulses. We find that pulse durations are limited to ps due to refraction and dispersion.

With current technology, timing ($\Delta t \sim 10^{-9}\,$s) is a better filter than frequency ($\sim 10^{-6}\,\Delta \lambda / \lambda_0$). This might be countered by new spectroscopic technologies on the receiver side, or power level advantages of continuous over pulsed lasers on the transmitter side.

The optimal laser signals to maximize $S/N$ appear to be time-bandwidth limited Gaussian $\Delta t \approx\,10^{-12}$\,s pulses at a wavelength $\lambda_{0}\approx1\,\mu$m, and a spectral width of $\Delta \lambda \approx 1.5\,$nm. An all-sky all the time survey at ps cadence may be performed given certain technological advances.

\acknowledgments
\textit{Acknowledgments}
MH is thankful to Marlin (Ben) Schuetz for useful discussions.

\end{document}